\documentclass[%
nofootinbib,
superscriptaddress,
preprintnumbers
]{revtex4}

\usepackage{%
amsmath,
amsthm,
amssymb,
braket,
bm,
ulem
}
\usepackage{graphicx}

\usepackage{color}

\begin{document}
\title{Limits on primordial magnetic fields from direct detection experiments of gravitational wave background}
\author{Shohei Saga}
\affiliation{Yukawa Institute for Theoretical Physics, Kyoto University, Kyoto 606-8502, Japan}
\author{Hiroyuki Tashiro}
\affiliation{Department of Physics and Astrophysics, Nagoya University, Nagoya, 464-8602, Japan}
\author{Shuichiro Yokoyama}
\affiliation{Kobayashi Maskawa Institute, Nagoya University, Aichi 464-8602, Japan}
\affiliation{Department of Physics, Rikkyo University, Tokyo 171-8501, Japan}
\affiliation{Kavli Institute for the Physics and Mathematics of the Universe (WPI),
Todai institute for Advanced Study, University of Tokyo, Kashiwa, Chiba 277-8568, Japan}

\date{\today}
\begin{abstract}

Primordial magnetic fields~(PMFs) can source gravitational wave
background~(GWB).
In this paper, we investigate
the possible constraints on small-scale PMF considering the ongoing and forthcoming
direct detection observations of GWB.
In contrast to the conventional cosmological probes, e.g.,~cosmic
microwave background anisotropies,
which are useful 
to investigate large-scale~PMFs~($>1~{\rm Mpc}$),
the direct detection experiments of GWB can explore small-scale PMFs whose scales correspond to the observed frequencies of GWB.
We show that 
future ground-based or space-based interferometric gravitational wave
detectors~give a strong constraint of about $10^{2}~{\rm nG}$ on much smaller scales of about $k\approx 10^{12}~{\rm Mpc}^{-1}$. 
We also demonstrate that pulsar timing arrays have a potential to strongly constrain PMFs.
The current limits on GWB from pulsar timing arrays can put the tight constraint on the amplitude of the PMFs of about $30~{\rm nG}$ whose coherent length is of about $k\approx 10^{6}~{\rm Mpc}^{-1}$.
The future experiments for the direct detection of GWB by the Square Kilometre Array could give much tighter constraints on the amplitude of PMFs about $5~{\rm nG}$ on $k\approx 10^{6}~{\rm Mpc}^{-1}$, on which scales, it is difficult to reach by using the cosmological observations.
\end{abstract}

\preprint{YITP-18-72, RUP-18-20}
\maketitle
\section{Introduction}
Recent direct detections of gravitational waves (GWs) from black hole
binary mergers and colliding neutron stars by LIGO/VIRGO collaboration
announce the coming of a new gravitational wave astronomy era
\cite{Abbott:2016blz,Abbott:2016nmj,Abbott:2017vtc,Abbott:2017oio,TheLIGOScientific:2017qsa,Abbott:2017gyy}. Obviously,
GWs from such astrophysical objects
give us valuable information about gravity in the strong field regime.
On the other hand, GWs from weak and unresolved sources constitute 
stochastic gravitational wave background~(GWB).
Although GWB still has not been detected, 
various experiments provide the upper limits on GWB in a wide range of frequencies~\cite{Arzoumanian:2015liz,Lentati:2015qwp,vanHaasteren:2011ni,Shannon:2013wma,TheLIGOScientific:2016dpb,2016PhLB..760..823P}.
There are many possible GWB source candidates proposed so far in both the standard cosmology and beyond.
Moreover, the evolution of GWB is sensitive to the expansion history of the Universe.
Therefore, the constraint on GWB is useful to reveal the physics of the early Universe, particularly, 
inflation models and the thermal history of the Universe (e.g.,~review by Refs.~\cite{Maggiore:1999vm,Romano:2016dpx}).

In the cosmological context, one of the important sources of GWB is the anisotropic stress of an energy
component of the Universe. 
Based on the cosmological perturbation theory in the linear regime,
any perturbations of the metric and the stress-energy tensor can be decomposed into
the scalar, vector, and tensor modes.
In the linear regime, they are decoupled in the Friedmann-Lema\^{\i}tre-Robertson-Walker (FLRW) universe.
Since the tensor mode of the anisotropic stress does not arise in the standard cosmology,
there is no GWB source after inflation.
However, in the nonlinear regime, these modes are coupled with each other, and
hence the GWs (corresponding to the tensor modes) can be sourced from
the anisotropic stress by the second-order terms of
the scalar and vector modes~\cite{Mollerach:2003nq,Ananda:2006af,Assadullahi:2009jc,Baumann:2007zm,Saga:2014jca}.
Therefore, from the limit on the GWB, we can obtain the constraint on the nature of the first-order scalar or vector modes.
For example, the current limits on the GWB by pulsar timing arrays~(PTAs)
put a constraint on the amplitude of the primordial density
fluctuations at small scales, and they impose a tight restriction on inflationary scenarios which could produce
a number of solar mass primordial black holes~\cite{Saito:2008jc,Nakama:2016gzw,Inomata:2016rbd,Ando:2017veq}.

If primordial magnetic fields (PMFs) exist in the expanding Universe, PMFs have an anisotropic stress, in particular, the tensor mode of the anisotropic stress, which can generate GWB.
The GWB generated from PMFs can affect the cosmic microwave
background~(CMB) temperature and polarization anisotropies at large scales~\cite{Durrer:1999bk,Mack:2001gc,Lewis:2004ef,Paoletti:2008ck,2010PhRvD..81d3517S}.
Therefore, recent CMB observations at large scales provide 
the upper limit on PMFs, which is in the order of nano-Gauss at Mpc scales.
PMFs can also affect CMB anisotropies directly through the magneto-hydro dynamics effects~\cite{2011arXiv1108.2517J,2013JCAP...10..050J}.
In particular, the stringent upper limit has been recently provided by Ref.~\cite{Jedamzik:2018itu}.
Using the numerical MHD simulations, they constrain
pico-Gauss magnetic fields on Mpc scales through the effect of PMFs on the recombination process.

In addition, the energy density or anisotropic stress of PMFs contribute to the primordial fluctuations as an isocurvature mode called a compensated magnetic mode~\cite{Shaw:2010ea,2010PhRvD..81d3517S,Zucca:2016iur}.
The effect of the compensated mode on the large-scale structure, i.e., matter power spectrum, appears at small scales.
From the observation of the large-scale structure of the Universe, we can
obtain the similar limit on PMFs as the upper bound from CMB anisotropies.

Cosmological observations can constrain PMFs on typically Mpc scales.
Since observations of small-scale PMFs can provide valuable information for
exploring the origin of cosmological magnetic fields,
many authors have conducted studies on the upper bound of PMFs at
smaller scales than Mpc
with various types of observations.
The constraint on the spectral distortion of CMB photons
can give a limit on PMFs of
several tens nano-Gauss at ${\rm kpc}$ scales due to the energy injection of decaying PMFs during the early stage of the Universe~\cite{Jedamzik:1999bm,Kunze:2013uja}.
The success of the big bang nucleosynthesis (BBN) can also provide the constraint on
the total energy of PMFs~\cite{2012PhRvD..86f3003K}. This constraint does not depend on the scale of PMFs.
The entropy production due to the energy dissipation of PMFs in the
early Universe can also give the limit on the PMFs at small scales, i.e.,~$k\gtrsim 10^{4}\; {\rm Mpc}^{-1}$~\cite{Saga:2017wwr}.
Note that before the recombination epoch, the nonlinear effect inevitably produces second-order magnetic fields as $10^{-24}$ Gauss in the standard cosmology, and therefore this value can be read as a ``theoretical'' lower bound on PMFs~\cite{Ichiki:2006cd,2011MNRAS.414.2354F,Saga:2015bna,Fidler:2015kkt}.

In this paper, we investigate the limits on the PMFs obtained from the direct
observations of GWB, for example, at pulsar timing arrays, (e.g.,~NANOGrav~\cite{Arzoumanian:2015liz}, European PTA~\cite{vanHaasteren:2011ni,Lentati:2015qwp}, and Parkes PTA~\cite{Shannon:2013wma}), at space-based GW
observatories, (e.g.,~LISA~\cite{AmaroSeoane:2012km}), and at ground-based GW observatories (e.g.,~LIGO~\cite{TheLIGOScientific:2016dpb}).
Although there is no report of the direct detection of GWB,
nondetection of GWB even in the current status of the observations
allows us to obtain a stringent constraint on the PMFs.
Since the direct measurements of GWB are sensitive to very high frequency GWB, in other words, very small scales,
these observations also give constraints on the PMFs with smaller scales, 
compared to the CMB measurement.

This paper is organized as follows. In the next section, we briefly review the GWB sourced by the anisotropic stress of the PMFs.
In Sec.~\ref{sec: result}, we present our main results and discussion.
First, we assume that the spectrum of PMFs is a delta-function type power spectrum, whose amplitude and characteristic scale are tightly constrained.
Next, we also explore the power-law type power spectrum, whose origin is assumed to be a cosmological phase transition.
In both cases, the direct observation of GWB can tightly constrain the amplitude of PMFs.
Finally, in Sec.~\ref{sec: summary}, we summarize this paper.

\section{Gravitational waves sourced from primordial magnetic fields}
In this section, we give the power spectrum of GWB sourced from PMFs, based on Refs.~\cite{Caprini:2001nb,2010PhRvD..81d3517S}.
If the PMFs are generated in the early Universe, they must induce an anisotropic stress in the energy-momentum tensor, which would be a source of the gravitational waves on both super- and sub-horizon scales.
By following Ref.~\cite{2010PhRvD..81d3517S}, the spatial $(i,j)$
components of the energy-momentum tensor for PMFs can be written in
terms of background pressure of photons~($\bar{p}_{\gamma}$), the density perturbation~($\Delta_{B}$), and anisotropic stress of PMFs~($\pi^{B}_{ij}$) as
\begin{eqnarray}
T^{i}{}_{j}(\eta, \bm{x}) &=& \frac{1}{4\pi a^{4}(\eta)}
\left( \frac{1}{2}B^{2}(\bm{x})\delta^{i}{}_{j} - B^{i}(\bm{x}) B_{j}(\bm{x})\right) ~, \\
&\equiv & \bar{p}_{\gamma}(\eta)\left( \Delta^{B}(\bm{x}) \delta^{i}{}_{j} + \pi^{B\,i}{}_{j} (\bm{x})\right) ~, \label{eq: def Pi}
\end{eqnarray}
where $\eta$ and $a(\eta)$ are the conformal time and scale factor, respectively.
Moreover, $\bm{B}(\bm{x})$ is the comoving magnetic fields,~i.e.,~$\bm{B}(\bm{x}) \equiv \bm{B}(\eta, \bm{x})/a^{2}$, where a factor $1/a^2$ comes from the adiabatic decay due to the cosmic expansion.

\begin{figure}[t]
\begin{center}
\includegraphics[width=0.5\textwidth]{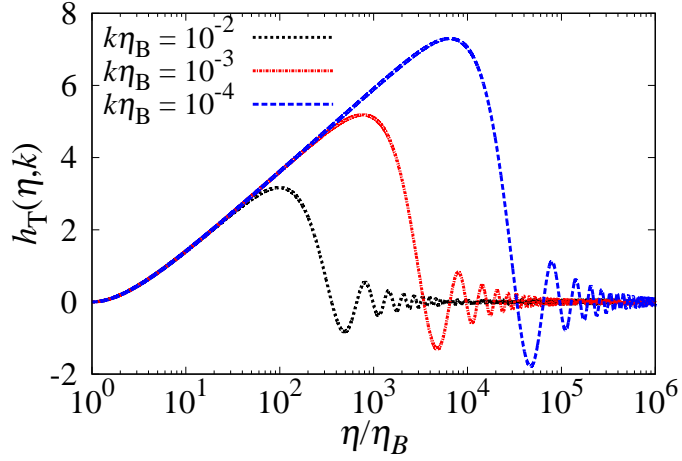}
\end{center}
\caption{The transfer function $h_{\rm T}(\eta, k)$ as a function of $\eta / \eta_{\rm B}$. 
}
\label{fig: evolve}
\end{figure}

Here we focus on the GWs and the perturbed metric around the FLRW universe can be taken as
\begin{equation}
{\rm d}s^{2} = a^{2}(\eta)\left[ - {\rm d}\eta^{2} + (\delta_{ij} + 2h_{ij}) {\rm d}x^{i} {\rm d}x^{j}\right] ~,
\end{equation}
where $h_{ij}$ is a transverse and traceless tensor perturbation.
From the Einstein equation with the energy momentum tensor given by Eq. (\ref{eq: def Pi}), the evolution equation for the Fourier component of $h_{ij}$ is given by
\begin{equation}
{h}_{ij}''(\eta, \bm{k}) + \frac{2}{\eta}{h}_{ij}'(\eta, \bm{k}) + k^{2} h_{ij}(\eta, \bm{k}) = \frac{1}{\eta^2} R_{\gamma} \pi^{B}_{ij}(\bm{k}) ~, \label{eq: hij}
\end{equation}
in the radiation-dominated era.
Here, a prime denotes the derivative with respect to the conformal time and $R_{\gamma} \equiv \bar{\rho}_{\gamma}/\bar{\rho}_{\rm r}$, where 
$\bar{\rho}_{\gamma}$ and
$\bar{\rho}_{\rm r}$ are respectively the energy densities of the photons and total radiation components.
The solution of Eq.~(\ref{eq: hij}) can be written as 
\begin{equation}
h_{ij}(\eta, \bm{k}) = h_{\rm T}(\eta, k)R_{\gamma} \pi^{B}_{ij}(\bm{k}) ~, \label{eq: h and pi}
\end{equation}
where $h_{\rm T}$ is a transfer function of GWs given by
\begin{equation}
h_{\rm T}(\eta, k) = -\frac{i}{2 k\eta}
\left[
e^{i k\eta} \left( {\rm Ei}(-i k\eta) - {\rm Ei}(-i k\eta_{\rm B}) \right)
- e^{-i k\eta} \left( {\rm Ei}(i k\eta) - {\rm Ei}(i k\eta_{\rm B}) \right)
\right] ~. \label{eq: transfer}
\end{equation}
Here, ${\rm Ei}(x)$ is the exponential integral, and $\eta_{\rm B}$
denotes an initial time which can be considered to be a generation time
of PMFs.
For the inflationary magnetogenesis scenarios, we assume $\eta_{\rm B}$ to be the beginning of the radiation
dominated era, that is, the reheating time.%
\footnote{
This assumption neglects the amplification during inflation, which might be strongly model dependent.
In that sense, our result shows the conservative upper bounds on PMFs.
}
As an initial condition, we take $h_{ij}(\eta_{\rm B}) = 0$ and $h_{ij}'(\eta_{\rm B}) = 0$.
We can find an approximate form of the above transfer function for the case with $k \eta_{\rm B} \ll 1$ as~\cite{2010PhRvD..81d3517S}
\begin{eqnarray}
h_{\rm T}(\eta, k) \approx \left\{ \begin{array}{ll}
\log \left( \frac{\eta}{\eta_{\rm B}} \right) + \frac{\eta_{\rm B}}{\eta} - 1 & ~ (k \eta \ll 1)~, \cr\cr
- \log (k \eta_{\rm B}) \, {\sin k\eta \over k\eta} & ~(k \eta \gg 1) ~.\\
\end{array} \right. 
\label{eq:approxtransfer}
\end{eqnarray}
We show the temporal evolution of the transfer function in Fig.~\ref{fig: evolve}.
The amplitude of GWB has a peak at the horizon crossing time, and after that, the amplitude decays as $h_{\rm T}(\eta,k) \propto \eta^{-1} \propto a^{-1}$ as shown in Eq.~(\ref{eq:approxtransfer}).

The power spectrum of GWs is defined as
\begin{equation}
\Braket{h_{ij}(\eta, \bm{k})h^{*}_{ij}(\eta, \bm{k'})} = (2\pi)^{3}\delta^{3}_{\rm D}(\bm{k}-\bm{k'})P_{h}(\eta, k) ~.
\end{equation}
The explicit form of the anisotropic stress of PMFs,~$\pi^{B}_{ij}$, is given by
\begin{equation}
\pi^{B}_{ij}(\bm{k}) = -\frac{3}{4\pi a^{4}\bar{\rho}_{\gamma}} \int{\frac{{\rm d}^{3}\bm{k}_{1}}{(2\pi)^{3}}}\; B_{i}(\bm{k}_{1}) B_{j}(\bm{k}-\bm{k}_{1}) ~.
\end{equation}
Therefore, by using Eq.~(\ref{eq: h and pi}),
we can evaluate the power spectrum of GWs sourced by PMFs as
\begin{eqnarray}
P_{h}(\eta, k) & =&
h^{2}_{\rm T}(\eta, k) R^{2}_{\gamma}\frac{3}{64\pi^{2}\rho_{\gamma,0}^{2}}
\int\frac{{\rm d}^{3}\bm{k}_{1}}{(2\pi)^{3}}
\int\frac{{\rm d}^{3}\bm{k}_{2}}{(2\pi)^{3}}
(2\pi)^{3}\delta^{3}_{\rm D}(\bm{k} - \bm{k}_{1} - \bm{k}_{2})
\notag \\
&&\times 
P_{B}(k_{1})P_{B}(k_{2})
\left( 1+ \left( \hat{\bm{k}}\cdot\hat{\bm{k}}_{1} \right)^{2} \right)
\left( 1+ \left( \hat{\bm{k}}\cdot\hat{\bm{k}}_{2} \right)^{2} \right) ~, \label{eq: power spectrum GW}
\end{eqnarray}
where the hat means the unit vector and $P_B(k)$ is a power spectrum of the PMFs.
Assuming that the PMFs are Gaussian and nonhelical,
the power spectrum of the PMFs can be written as
\begin{equation}
\Braket{B_{i}(\bm{k}) B^{*}_{j}(\bm{k'})} = \frac{(2\pi)^{3}}{2}\delta^{3}_{\rm D}(\bm{k}-\bm{k'}) \left( \delta_{ij} - \hat{k}_{i}\hat{k}_{j}\right)P_{B}(k) ~.
\end{equation}
Finally, we can calculate the density parameter of GWB from PMFs with the power
spectrum in Eq.~(\ref{eq: power spectrum GW})
by
\begin{equation}
\Omega_{\rm GW}(\eta, k) = \frac{1}{12}\left( \frac{k}{aH}\right)^{2} \frac{k^{3}}{2\pi^{2}} P_{h}(\eta, k) ~.
\end{equation}

\section{Results and Discussions}\label{sec: result}
In this section, we discuss the upper bound of PMFs through the measurement in the direct detection experiments of GWB.
For the simplicity of analysis, first the power spectrum of PMFs is assumed to be the delta-function type as
\begin{equation}
P_{B}(\ln{k}) = \frac{2\pi^{2}}{k^{3}}\mathcal{B}^{2}\delta_{\rm D}\left( \ln{\left(k/k_{\rm p}\right)} \right) ~.
\label{eq: delta power}
\end{equation}
For this delta-function type 
of PMFs, the energy density of GWB at the present time ($\eta =
\eta_{0}$) can be represented as
\begin{equation}
\Omega_{\rm GW}(\eta_{0}, k) = \frac{R^{2}_{\gamma}}{512\pi^{2}} \left( \frac{\mathcal{B}^{2}}{\bar{\rho}_{\gamma,0}} \right)^{2}
\left( \frac{k}{H_{0}} \right)^{2} a^{2}_{\rm eq} h^{2}_{\rm T}(k,\eta_{\rm eq}) \left( \frac{k}{k_{\rm p}}\right)^{2} \left( 1 + \frac{k^{2}}{4k^{2}_{p}}\right)^{2} \Theta_{\rm H}\left( 1-\frac{k}{2k_{\rm p}} \right) ~, \label{eq: delta OmegaGW}
\end{equation}
where $\Theta_{\rm H}(x)$ is the Heaviside step
function and 
the subscript ``eq'' means the value at the epoch of
matter-radiation equality.
Since we are interested in GWB whose wavelengths are much smaller than the horizon scale at $\eta_{\rm eq}$,
we simply adopt the adiabatic
evolution after the epoch of matter-radiation equality in order to
obtain Eq.~\eqref{eq: delta OmegaGW}.
That is,
the amplitude of GWB at the present epoch~$\eta_0$, $h_{\rm T}(k,\eta_{0})$
can be given by $h_{\rm T}(k,\eta_{0})a_{0} = h_{\rm T}(k,\eta_{\rm
eq})a_{\rm eq}$.

Note that, in the above analysis, we assume that the anisotropic stress of
neutrinos can be neglected. This assumption can be justified as follows.
After the neutrino decoupling era, neutrinos start to stream freely, and
the additional contribution appears in the rhs in Eq.~(\ref{eq: hij})
as the anisotropic stress of neutrinos.
The neutrino anisotropic stress should be described as~\cite{Weinberg:2003ur,Watanabe:2006qe}
\begin{equation}
R_{\nu}\pi^{\nu}_{ij}(\eta, k) = -24R_{\nu}\int^{\eta}_{\eta_{\nu}}{\rm d}\eta_{1}\; \frac{j_{2}\left( k(\eta - \eta_{1})\right)}{k^{2}(\eta -\eta_{1})^{2}} \dot{h}_{ij}(\eta, k) ~,
\end{equation}
where $j_{2}(x)$ is a spherical Bessel function and $R_{\nu} = \bar{\rho}_{\nu}/\bar{\rho}_{\rm r}$.
From the above expression, one can find that 
the effect of the neutrino anisotropic stress on the evolution of GWs
would be negligible on subhorizon scales even after the neutrino decoupling ($k \eta \gg 1$, $\eta > \eta_\nu \approx 7.6\times 10^{-4}~ {\rm Mpc}$).
The observations of PTAs, which are the current lowest frequency experiments
for direct detection, can be sensitive to GWB with $k_{\rm PTA}\approx 5\times 10^{6}\; {\rm Mpc}^{-1}$.
Therefore, as long as we consider scales larger than the direct GW observations,
e.g.,~PTAs and GW interferometers,
we can safely neglect the effect of neutrino anisotropic stress.

Now we evaluate Eq.~\eqref{eq: delta OmegaGW} numerically.
Before we move on,
it is helpful to 
remove the oscillation part from the transfer function of Eq.~(\ref{eq:
transfer}) for numerical evaluation.
Therefore, we approximate Eq.~(\ref{eq: transfer}) to
\begin{equation}
h^{2}_{\rm T}(\eta_{\rm eq},k) \approx 
\frac{ \left( {\rm Ci}(k\eta_{\rm eq})-{\rm Ci}(k\eta_{B}) \right)^{2} + \left( {\rm Si}(k\eta_{\rm eq})-{\rm Si}(k\eta_{B}) \right)^{2} }{(k\eta_{\rm eq})^{2}} ~, \label{eq: shape}
\end{equation}
where ${\rm Ci}(x)$ and ${\rm Si}(x)$ are the cosine-integral and sine-integral, respectively.
This approximation is valid only for $\eta^{-1}_{\nu} \lesssim k$.
In the case of the direct detection of GWB,
the condition, $\eta^{-1}_{\nu} \lesssim k$, is well satisfied as we have mentioned.

\begin{figure}[t]
\begin{center}
\includegraphics[width=0.5\textwidth]{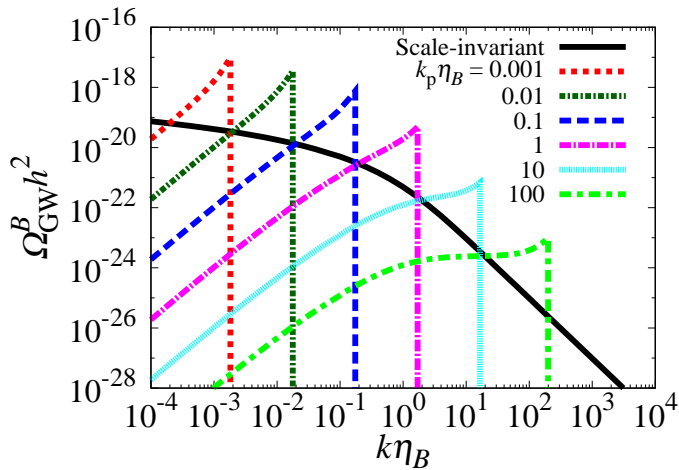}
\end{center}
\caption{The spectrum of GWB induced by the anisotropic stress of PMFs at the present time for various $k_{\rm p}$.
In this figure, we set $\mathcal{B} = 1\; {\rm nG}$.
The amplitude of GWB is scaled proportional to $\mathcal{B}^{4}$.
We also show the scale-invariant case defined in Eq.~(\ref{eq: scale-inv}) with the solid black line.
}
\label{fig: spectrum}
\end{figure}

Plugging Eq.~\eqref{eq: shape} into Eq.~\eqref{eq: delta OmegaGW},
we calculate the energy density of GWB at the present time.
Figure \ref{fig: spectrum} represents the results, $\Omega_{\rm GW}$,
as a function of $k$.
Here we set ${\cal B}=1~$nG.
In Fig.~\ref{fig: spectrum},
we also show the dependence on $k_{\rm p} \eta_{B}$, taking different $k_{\rm p} \eta_{B}$ from $0.001$ to $100$.
In this figure, although we set $\eta_B/\eta_{\nu} = 10^{-12}$, we confirm that the spectra are insensitive to the choice of $\eta_{B}$.
%
As can be seen in this figure, the induced GWB has a peak at $k = k_{\rm p}$ and
the peak amplitude of GWB is almost saturated for $k_{\rm p} \eta_{\rm B} \ll 1$.
This is because the amplitude of GWB at $k = k_{\rm p}$ depends on $k_{\rm p}\eta_B$ logarithmically for the case with $k \eta_{\rm B} \ll 1$, as shown in Eq.~\eqref{eq:approxtransfer}. 
On the other hand, the amplitude of generated GWB is strongly suppressed on $k_{\rm p}\eta_{B} \gg 1$.
For this reason,  in the case of relatively higher frequency experiments such as LIGO
where the observed frequency, $k_{\rm obs}$, can become larger than $1/\eta_{\rm B}$, the constraints would strongly depend on the generation time, $\eta_{\rm B}$.
Note that a causal generation can produce only
magnetic fields whose scale is less than the horizon scale at the
generation epoch.
However, causal magnetic fields can have the power
even on superhorizon scales in the Fourier space
because causal magnetic fields have the tail of the blue power spectrum on
scales larger than the horizon scale~(see Ref.~\cite{Durrer:2003ja}).
Therefore, considering the above power spectrum (\ref{eq: delta power}) allow us to obtain 
the constraint on the small-$k$ tail part of causal magnetic fields.

Here, we consider three types of observations, i.e., PTAs, space-based GW
observatories, and ground-based GW observatories.
First, PTAs can be sensitive to
GWB with $k_{\rm PTA}\approx 5\times 10^{6}\; {\rm Mpc}^{-1}$, e.g.,~\cite{1979ApJ...234.1100D,Hellings:1983fr}.
We refer to the results for the current running PTAs as NANOGrav~\cite{Arzoumanian:2015liz}, European PTA~\cite{vanHaasteren:2011ni,Lentati:2015qwp}, and Parkes PTA~\cite{Shannon:2013wma}, and future PTA project as Square Kilometre Array (SKA)~\cite{Janssen:2014dka}.
Second, 
for a space-based GW observatory
we consider LISA which is planned now.
In the current design of LISA,
it is expected that GWB could be strongly constrained~\cite{AmaroSeoane:2012km}.
Third, as the upper bound for a ground-based GW observatory, we adopt the recent report by LIGO~\cite{TheLIGOScientific:2016dpb}.
We summarize these (expected) upper bounds in Table.~\ref{tab: observations} with the most sensitive wave numbers and corresponding upper limits.
\begin{table}[t]
\begin{tabular}{c||c|c}
& wave number $k$ [${\rm Mpc}^{-1}$] & Upper limit on $\Omega_{\rm GW}h^{2}$ \\
\hline\hline
Current PTAs~\cite{vanHaasteren:2011ni,Lentati:2015qwp,Shannon:2013wma,Arzoumanian:2015liz} & $\approx 5\times 10^{6}$ & $\lesssim10^{-9}$ \\
\hline
LIGO~\cite{TheLIGOScientific:2016dpb} & $\approx 10^{17}$ & $\lesssim10^{-7}$ \\
\hline
SKA~\cite{Janssen:2014dka} & $\approx 5\times 10^{6}$ & $\lesssim10^{-13}$ \\
\hline
LISA~\cite{AmaroSeoane:2012km} & $\approx 10^{12}$ & $\lesssim10^{-9}$ \\
\hline
\end{tabular}
\caption{Summary of the observations we assumed. Current PTA and LIGO bounds are obtained from the observed results but SKA and LISA are expected upper bounds in the future.}
\label{tab: observations}
\end{table}

\begin{figure}[t]
\begin{center}
\includegraphics[width=0.5\textwidth]{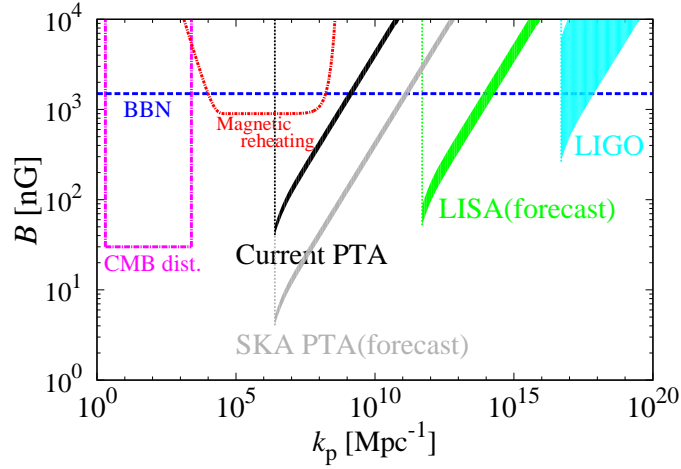}
\end{center}
\caption{
Upper bounds on the amplitude of PMFs obtained from direct detection measurements of GWB; current PTAs~(black shaded), SKA PTA~(gray shaded), LISA~(green shaded), and LIGO~(cyan shaded).
We also show the previous studies: magnetic reheating~(red)~\cite{Saga:2017wwr}, BBN (blue)~\citep{2012PhRvD..86f3003K} and CMB distortion (magenta)~\citep{Jedamzik:1999bm}.
Upper bounds from the direct detection measurements of GWB are shown by the shaded regions which come from the generation epoch of PMFs within $10^{-17} \leq \eta_{B}/\eta_{\nu} \leq 10^{-12}$.
}
\label{fig: limit}
\end{figure}

We summarize our constraints in Fig.~\ref{fig: limit}.
The upper bounds estimated from the direct detection experiments of GWB are expressed in solid lines.
The thickness of lines corresponds to the range of the PMF generation epoch and we take it to be $10^{-17} \leq \eta_{B}/\eta_{\nu} \leq 10^{-12}$.
The bottom lines correspond to the upper bound for the case with $\eta_{B}/\eta_{\nu} = 10^{-17}$.
As we have mentioned, for the experiments with relatively higher frequency bands such as LIGO, 
the amplitude of generating GWB strongly depends on the PMF generation epoch, $\eta_B$, and hence
the solid line for LIGO seems to be thicker. As a result, current PTAs and LIGO give $\mathcal{B} \lesssim 40~{\rm nG}$ for $k \approx 10^6~{\rm Mpc}^{-1}$ and $\mathcal{B} \lesssim 300~{\rm nG}$ for $k \approx 10^{17}~{\rm Mpc}^{-1}$ (for $\eta_B / \eta_\nu = 10^{-17}$), respectively.
LISA is expected to give $\mathcal{B} \lesssim 50~{\rm nG}$ for $k
\approx 10^{12}~{\rm Mpc}^{-1}$. The PTA by SKA will give a tight constraint on the amplitude of PMFs as $\mathcal{B} \lesssim 4~{\rm nG}$ for $k \approx 10^{6}~{\rm Mpc}^{-1}$. 

So far we have considered the delta-function type of the PMF power spectrum to make it easy to understand the correspondence between the scale of PMFs and the frequency of induced GWB.
Let us consider the PMFs with the power-law
spectrum as a more general case.
First, we consider the scale-invariant spectrum whose form is assumed to be
\begin{equation}
P_{B}(k) = \frac{2\pi^{2}}{k^{3}} \mathcal{B}^{2}\times
\begin{cases}
\ln{\left( k_{\rm max}/k_{\rm min}\right)}^{-1} & k_{\rm min}\leq k\leq k_{\rm max}\\
0 & \mbox{otherwise}
\end{cases}~,
\label{eq: scale-inv}
\end{equation}
where we introduce IR and UV cutoffs with $k_{\rm min} = \eta^{-1}_{\nu}$ and $k_{\rm max}\eta_{B} =
10^{8}$.
Note that we have confirmed that the result does not depend on the choice of these cutoff 
scales.
In Fig.~\ref{fig: spectrum}, we plot $\Omega_{\rm GW}$ due to the
scale-invariant spectrum as a solid black line.
As seen in the case of the delta-function type,
the scale dependence of the energy density of induced GWB becomes weaker at larger scales than the horizon scale at the generation epoch.
For the scale-invariant case where the power spectrum is assumed to be Eq. \eqref{eq: scale-inv}, the current PTAs give an upper bound as
$\mathcal{B} \lesssim 2.5 \times 10^{2}\; {\rm nG}$.

Finally, we briefly mention the case where the PMFs are causally generated~\cite{Durrer:2003ja}
and generated PMFs have a blue-tilted power spectrum assumed to be~\cite{Ade:2015cva}
\begin{equation}
P_{B}(k) = \frac{2\pi^{2}}{k^{3}}
\frac{2 (2\pi)^{n_{B}+3} B^{2}_{\lambda}}{\Gamma\left( \frac{n_{B}+3}{2}\right)}
\left( \frac{k}{k_{\lambda}} \right)^{n_{B}+3} \Theta_{\rm H}( k_{\rm c} - k) ~,
\end{equation}
where we introduce the Heaviside step function $\Theta_{\rm H}(x)$ which means that the amplitude of PMFs is identical to zero at smaller scales than the cutoff scale $k_{\rm c}$.
Here $B_{\lambda}$ is the amplitude of PMFs by smoothing over comoving scale of $\lambda$ and $k_{\lambda} \equiv 2\pi/\lambda$.
For such blue-tilted PMFs, the spectrum of the energy density of GWB is given as
\begin{eqnarray}
\Omega_{\rm GW}(k,\eta_{0})
&=&
\frac{R^{2}_{\gamma}}{512\pi^{2}}
\left( \frac{2 (2\pi)^{n_{B}+3}}{\Gamma\left( \frac{n_{B}+3}{2}\right)} \right)^{2}
\left( \frac{B_{\lambda}^{2}}{\rho_{\gamma,0}} \right)^{2}
\left( \frac{k}{H_{0}}\right)^{2} h^{2}_{\rm T}(k,\eta_{0})
\left(\frac{k}{k_{\lambda}}\right)^{2}
\int^{\infty}_{0} \frac{{\rm d}k_{1}}{k_{1}}\; 
\left( \frac{k_{1}}{k_{\lambda}}\right)^{n_{B}+2}
\Theta_{\rm H}(k_{\rm c}-k_{1})
\notag \\
&&\times 
\int^{k+k_{1}}_{|k-k_{1}|} \frac{{\rm d}k_{2}}{k_{2}}\;
\left( \frac{k_{2}}{k_{\lambda}}\right)^{n_{B}+2}
\Theta_{\rm H}(k_{\rm c} - k_{2})
\left( 1+ \left( \hat{\bm{k}}\cdot\hat{\bm{k}}_{1} \right)^{2} \right)
\left( 1+ \left( \hat{\bm{k}}\cdot\hat{\bm{k}}_{2} \right)^{2} \right) ~.
\end{eqnarray}
If we assume $k/k_{1} \ll1$ and $2n_{B}+3 > 0$, we can perform the integrations in terms of $k_{2}$ and $k_{1}$ and obtain an approximate expression as
\begin{eqnarray}
\Omega_{\rm GW}(k,\eta_{0})
&\approx&
\frac{R^{2}_{\gamma}}{64\pi^{2}}
\frac{(2\pi)^{2n_{B}+6}}{\left[ \Gamma\left( \frac{n_{B}+3}{2}\right) \right]^{2}}
\frac{1}{2n_{B}+3} \left( \frac{k_{\rm c}}{k_{\lambda}}\right)^{2n_{B}+3}
\left( \frac{B_{\lambda}^{2}}{\rho_{\gamma,0}} \right)^{2}
\left( \frac{k}{H_{0}}\right)^{2}
\left(\frac{k}{k_{\lambda}}\right)^{3}
h^{2}_{\rm T}(k,\eta_{0}) ~~~ \mbox{(for $k<k_{\rm c}$)}~,
\end{eqnarray}
From the above approximate expression, 
one can find that the scale dependence of $\Omega_{\rm GW}$ is independent of the spectral index of PMFs $n_{B}$, i.e., proportional to $k^{2}h^{2}_{\rm T}(k,\eta_{0})$,
while the amplitude of $\Omega_{\rm GW}$ depends on it~\cite{Durrer:1999bk,Caprini:2001nb}.
As an example, let us assume the PMFs generated at the electroweak phase transition where $\eta_{B}$ is taken to be equal to $\eta_{\rm EW} \sim 10^{-6} \eta_\nu$ and the spectral index of PMFs is expected to be $n_{B} = 2$~\cite{Durrer:2003ja}.
For such a case, the cutoff scale is assigned to the horizon-scale of the electro-weak transition, i.e. $k_{\rm c} = \eta^{-1}_{\rm EW}$.
Therefore, since the observed frequency band of LIGO is much larger than the cut-off scale in the spectrum of PMFs, $k_{\rm c}$, the PTA observations can put a strong constraint on the amplitude of PMFs.

The current PTA observations put the constraint on the amplitude of
PMFs as $B_{1\,{\rm Mpc}} \lesssim 1.9\times 10^{-18}\; {\rm nG}$.
This constraint is comparable to that obtained from the nucleosynthesis bound on GWB~\cite{Caprini:2001nb}.
The future PTA observation by SKA is expected to constrain the amplitude of
PMFs about $B_{1\,{\rm Mpc}} \lesssim 1.0\times 10^{-19}\; {\rm nG}$. The expected constraint by the future LISA experiments has been investigated in~\cite{Caprini:2001nb,Caprini:2009yp}.
In the above analysis, we focus on PMFs generated in the cosmological phase transition.
However, Ref.~\cite{Anand:2018mgf} discusses the upper bound of GWB due to turbulence in the chiral plasma sourced by PMFs.
Even in this specific model, GWB induced from PMFs can be also strongly constrained by the PTA.
In particular, the future PTA observation such as SKA should be a good
probe to explore various models of the PMF generation.

\section{Summary}\label{sec: summary}
Under the presence of primordial magnetic fields (PMFs), the tensor mode in the anisotropic stress of PMFs can generate gravitational wave background (GWB).
Although the PMFs at large scales are well constrained by the cosmological probes such as the CMB anisotropies, small-scale PMFs are less done.
In this paper, we establish the upper limit on the PMFs through various experiments on the direct detections of GWB.
The sensible scales for the direct detection of GWB are widely broadened from $k \approx 10^{6} \sim 10^{17} \; {\rm Mpc}^{-1}$, and therefore the limit on PMFs at similar scales can be obtained.
In this sense, the direct detection of GWB is one of the keys to explore the signature at small scales.

The nature of PMFs can be described by the primordial power spectrum of PMFs in which the generation mechanism would be imprinted.
First, we assume the delta-function type power spectrum with two PMF
parameters, its amplitude and the scale of the peak position.
We find that the PTA can strongly constrain the amplitude of PMFs on scales $k\approx 10^{6}\sim10^{9}\; {\rm Mpc}^{-1}$.
In particular, the future observations of PTAs such as the Square Kilometre Array has a potential to put a limit on PMFs at about $5\;{\rm nG}$ at $k\approx 10^{6} \; {\rm Mpc}^{-1}$, which we cannot access by using the conventional cosmological observations.
We should note that, although the magnetic reheating or BBN can also constrain similar scales, those upper bounds are weaker than that from the PTAs.
We also study the case of the power-law type power spectrum, especially the scale-invariant power spectrum.
Even in this case, the direct observations by PTAs such as SKA are also better probes for putting an upper limit on the amplitude of PMFs.
In particular, the current PTAs give an upper bound as $2.5\times 10^{2} \; {\rm nG}$.
Finally, when the origin of the PMFs is assigned to the cosmological phase transitions, the spectrum of PMFs can be described as a power-law type power spectrum with a blue power tilt.
Considering the causal PMF generation, the tilt of the power spectrum of PMFs can be set as $n_{B} = 2$.
In this case, the amplitude of PMFs with smoothing over a comoving scale of $\lambda = 1\; {\rm Mpc}$ is bounded at about $10^{-18}\; {\rm nG}$ from current PTA observations.
This upper limit will be comparable to the limit from future experiments such as LISA, except for the constrained scales.
Moreover, future PTA observations such as SKA can put a stronger constraint on the PMFs of  $10^{-19}\; {\rm nG}$.

We can conclude that, in either case, the direct observations of GWB from PTAs work well in order to constrain PMFs.
In particular, SKA would be a promising probe for accessing the small-scale PMFs that cosmological observations cannot reach.
Note that throughout this paper, we focus only on nonhelical PMFs.
However, if we add the helical components of PMFs, the helical GWB emerges, and moreover, nonhelical GWB is also amplified by the helical PMFs~\cite{Durrer:1999bk}.
These objects will be presented in a future work.

\acknowledgements
This work is supported in part by a Grant-in-Aid for Japan Society for Promotion of Science (JSPS) Research Fellow Number 17J10553~(S.S.), JSPS KAKENHI Grant Number 15K17646~(H.T.), 17H01110~(H.T.) and 15K17659~(S.Y.), and MEXT KAKENHI Grant Number 18H04356~(S.Y.).
\bibliography{ref}
\end{document}